\documentstyle[prb,epsfig,aps]{revtex}

\input epsf

\newcommand{\centps}[2]{
        \begin{center}
                \epsfig{file=#1,height=#2mm}
        \end{center}
}

\newcommand{\twofig}[3]{
        \begin{center}
                \epsfig{file=#1,height=#3truecm}
                \epsfig{file=#2,height=#3truecm}
        \end{center}
}

\begin{document}
\draft

\twocolumn[\hsize\textwidth\columnwidth\hsize\csname @twocolumnfalse\endcsname



\widetext

\title{Excess Currents Larger than the Point Contact Limit
in Normal Metal -Superconducting Junctions}
  
\author{Richard Riedel and Philip F. Bagwell \\  
Purdue University\\  
School of Electrical Engineering and Computer Science\\  
West Lafayette, Indiana 47907}  

\date{\today}
\maketitle

\maketitle  

\widetext 

\begin{abstract}

In a point contact NS junction, perfect Andreev reflection occurs over
a range of voltages equal to the superconducting energy gap, producing
an excess current of $I_{\rm exc} = (4/3)(2e\Delta/h)$.  If the
superconductor has a finite width, rather than the infinite width of
the point contact, one cannot neglect superfluid flow inside the
superconducting contact. The energy range available for perfect
Andreev reflections then becomes larger than the superconducting
gap, since superfluid flow alters the dispersion relation inside the
finite width superconductor. We find a maximum excess current of
approximately $(7/3)(2e\Delta/h)$ when the width of the superconductor
is approximately $7/3$ times the width of the normal metal. 

\end{abstract}

\vspace{0.20in}

] \narrowtext


\section{Introduction}

\label{secintro}

\indent  

At a normal metal - superconductor (NS) junction, electrons incident
from the normal metal can be scattered into time reversed electrons
(holes) by the pairing potential.  This conversion process is known as
Andreev reflection.~\cite{rAndreev} When the NS junction carries an
electrical current, Andreev reflection is accompanied by the
conversion of normal current to supercurrent.~\cite{rKum,rMathews}
This supercurrent flow modifies the current-voltage relation in NS and
NSN junctions~\cite{rSanSol}~\cite{rSanSol2}~\cite{rLamb},
superconducting wires~\cite{rLiFu,rBag1}, SNS junctions~\cite{rRie1},
and NS junctions with a supercurrent parallel to the NS
interface~\cite{rHofKum}.  In this paper we consider the
current-voltage relation for NS junctions having a supercurrent flow
perpendicular to the NS junction.

\begin{figure}[htb]
\centps{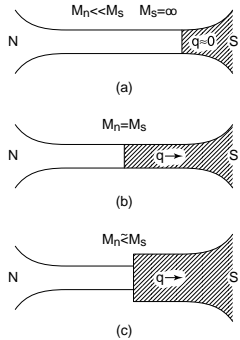}{45}
\caption{Different types of NS Junctions.  (a) NS point contact, (b)
N-Narrow S- S, (c) N - Wider S- S. The wire width determines the
number of conducting modes in the narrow segments ($M_N$ and
$M_S$). The superfluid flow velocity $v_s$ cannot be neglected when
$M_N \simeq M_S$.}
\label{fig1DGeom}
\end{figure} 

The superfluid flow present for the point contact NS
junction in Fig.~\ref{fig1DGeom}(a) will have little effect on its I-V
characteristic, since the current density inside the wide
superconductor approaches zero. However, for the NS junctions shown in
Fig.~\ref{fig1DGeom}(b)-(c) , the number of conducting modes in the
superconducting wire ($M_S$) is comparable to to the number of
conducting modes in the normal wire ($M_N$).  Since the current
density is not zero inside the superconductor, one cannot neglect the
effect of a superfluid flow on the I-V characteristics of the NS
junctions shown in Fig.~\ref{fig1DGeom}(b)-(c).  Since the superfluid
flow strongly modifies the dispersion relationship of the
superconductor when the ratio of the number of conducting modes
$\alpha = M_S / M_N$ is of order one, including such a superfluid flow
will influence the I-V characteristic of NS junctions. Since the
number of conducting modes is roughly proportional to the width of the
conductor, namely $\alpha \simeq W_S / W_N$, we can vary the ratio of
conducting modes by varying the width of the superconductor $W_S$.

Blonder, Tinkham and Klapwijk (BTK) showed that the point contact NS
junction in Fig.~\ref{fig1DGeom}(a) carries a larger current than a
normal metal point contact junction~\cite{rBTK}.  This 'excess'
current of is due to the presence of Andreev reflection, and has the
value $I_{\rm exc} = (4/3)(2e\Delta/h)$ in a ballistic point-contact
NS junction.  By varying the width of the superconducting wire forming
the NS junction in Fig.~\ref{fig1DGeom}(c), we determine in this paper
how the excess current varies as a function of the transverse mode
ratio $\alpha$. Using both an independent band model, and a second
model which includes scattering between different lateral modes at the
NS interface, we show the excess current in NS junction can be much
larger than the BTK result. This enhancement of the excess current over
the BTK value has also been noted in Ref.~\cite{rSanSol2}.

More Andreev reflections can occur at higher voltages when a
supercurrent flows perpendicular to the NS interface, accounting for
an excess current larger than the BTK result. If the superconductor is
narrower than about $W_S < (7/3) W_N$, the narrow region of the
superconductor exceeds its critical current before allowing the
maximum number of Andreev reflections. If the superconductor is much
wider than $W_S >> (7/3) W_N$, there is too much geometrical dilution
of the supercurrent for a significant superfluid flow to develop
inside the narrowest region of the superconductor. We find a maximum
excess current when the width of the superconductor is approximately
$W_S = (7/3) W_N$.

\section{Why Superfluid Flow Increases the Excess Current}
\label{secwhymore}

\indent

In this section we give the simplest physical model which illustrates
why the excess current in NS junctions can be larger than the point
contact limit. To make our physical points, we construct a crude two
fluid model, which does not obey electrical current conservation at
every point in space.  A fully self-consistent solution of the
Bogoliubov-deGennes equations automatically ensures current
conservation, eliminating the need for this ad-hoc two-fluid model,
but requires more computational effort. The two-fluid model we develop
only guarentees current conservation at the terminal contacts.  A
fully self-consistent solution of the BdG equations in a single mode
junction ($\alpha = 1$), done in section \ref{secself1d}, gives
the same results for the excess current as the two-fluid
model.  Viewing the electrical conduction in terms of this two fluid
model, therefore, allows us to obtain a value of the superfluid
velocity $v_s$ inside the superconducting contacts for each value of
the bias voltage across the NS or NIS junction. We then use this value
of the flow velocity $v_s$ to compute Andreev and normal reflection
probabilities for each value of the voltage, and thus obtain the
electrical current in a globally self-consistent manner.

\subsection{Two Fluid Model}
\label{twoflu}

\indent

\begin{figure}[htb]
\centps{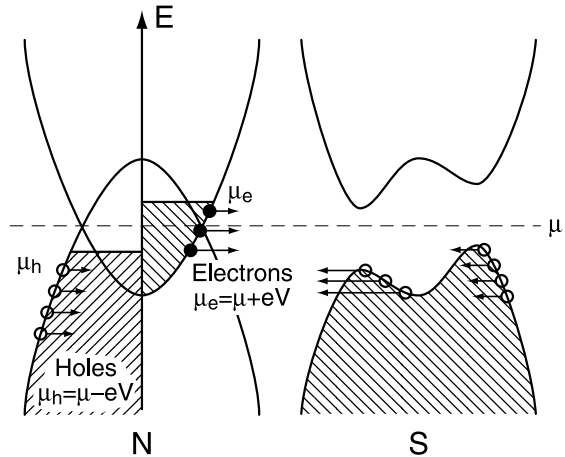}{60}
\caption{ Energy band diagram of an NS junction subject to a
superfluid flow $v_S \ge 0$.  The shifted energy bands in the
superconductor cause Andreev reflection to occur at higher energies
than without superfluid flow. The contacts inject electron-like (solid
dots) and hole-like (open circles) quasi-particles as shown.}

\label{figBand}
\end{figure}

Figure~\ref{figBand} shows the energy band diagram of an NS junction
when the superconductor carries a finite supercurrent~\cite{rBag1}.
In a transmission formalism, one must compute the electrical current
operator for all incident quasi-particle states, and add them to
obtain the total current. Ref.~\cite{rDatta1} evaluates the electrical
current operator on the normal side of the NS junction in terms of
particle current transmission and reflection probabilities.  The
derivation in section 3 of Ref.~\cite{rDatta1} is valid for a
multiple moded NS junction subject to a superfluid flow.
Ref.~\cite{rDatta1} recovers the well-known BTK current 
formula~\cite{rBTK}, namely (for zero temperature)
\begin{equation}  
I  = (2e/h) M_N \int^{eV}_{0} 
[1 - T_{Ne,Ne}(E) + T_{Nh,Ne}(E)] dE.
\label{I_NT=0}
\end{equation}
In Eq.~(\ref{I_NT=0}) we use the notation of Ref.~\cite{rDatta1},
where the $T_{N \delta, N \beta}$ particle current reflection
probabilities from the incident channel $(N \beta)$ to reflected
channel $(N \delta)$.  The indices $\beta, \delta = e \; {\rm or } \;
h$ for electron-like or hole-like quasi-particles.  We must compute
the both the normal $T_{Ne,Ne}=R_N$ and Andreev $T_{Nh,Ne}=R_A$
reflection probabilities when the superconducting contact is subject
to a superfluid flow.  For simplicity we have taken the $M_N$ modes in
the normal conductor to be both independent and to carry identical
electrical currents.

In order for the superconductor to carry a finite supercurrent, the
form of the order parameter $\Delta(x)$ inside the superconductor must
be generalized to $\Delta(x) = |\Delta| e^{2iqx}$. The superfluid
velocity is $v_{s}= \hbar q / m$. 
If we consider solutions accurate within the Andreev approximation
($|E| \ll \mu$), we can approximate the dispersion relation near the
Fermi level as being rigidly shifted in energy as~\cite{rBag1}
\begin{eqnarray}  
\frac{\hbar^{2}k^{2}}{2m} - \mu 
\simeq \pm \sqrt{(E \mp |\Delta| (q/q_d))^{2} -  |\Delta|^{2}} \; .
\label{shiftE}
\end{eqnarray}  
Here $v_d = \hbar q_d / m = |\Delta| / \hbar k_F$ is the Landau
depairing velocity.~\cite{rLan} The electrical current is given by
$I_C \simeq e n v_s M_S$ with $n = 2(k_F/\pi)$ the electron density
per mode and $M_S$ the number of (equivalent, for simplicity)
conducting modes in the superconductor~\cite{rLiFu}.  We can rewrite
the supercurrent $I_C$ in terms the depairing velocity as
\begin{equation}
I_C \simeq (4e|\Delta|/h) (v_s/v_d) M_S .
\label{I_C}
\end{equation}
At zero temperature, the critical current phase boundary occurs when
$v_s = v_d$.  We must therefore maintain $v_s < v_d$ to preserve the
superconducting order parameter $|\Delta| \ne 0$.

Inside the superconductor, the electrical current
is often argued to be composed of a `quasi-particle' and `condensate'
contribution~\cite{rBTK,rPethik}. To break Eq.~(\ref{I_NT=0}) down
into these two contributions to the current we use the sum
rule~\cite{rBTK}
\begin{equation}  
1 = T_{Ne,Ne}(E) + T_{Nh,Ne}(E) + T_{Se,Ne}(E)  + T_{Sh,Ne}(E).
\label{Pconserv}
\end{equation}  
Equation~(\ref{Pconserv}) states that the normal ($T_{Se,Ne}=T_{N}$)
and Andreev ($T_{Sh,Ne}=T_{A}$) particle current transmission
coefficients into the superconductor conserve the total number of
quasi-particles.  Combining Eqs.~(\ref{I_NT=0}) and (\ref{Pconserv}),
the electrical current inside the superconductor at zero temperature
is
\begin{equation}
I = I_{QP} + I_A.
\label{I_S}
\end{equation}  
We identify the portion of the electrical current due to
`quasi-particle' injection as
\begin{equation}  
I_{QP} = (2e/h) M_N \int^{eV}_{0} 
[T_{Se,Ne}(E) - T_{Sh,Ne}(E)] dE,
\label{I_QP}
\end{equation}
and the `Andreev' portion of the current $I_A$ as
\begin{equation}
I_A = (2e/h) M_N \int^{eV}_{0} 2 
[T_{Nh,Ne}(E)+ T_{Sh,Ne}(E) ] dE .
\label{2And}
\end{equation}
The `Andreev' current in Eq.~(\ref{2And}) equals
twice the sum of all Andreev processes. 

Equations~(\ref{I_S})-(\ref{2And}) give the same current as the BTK
expression from Eq.~(\ref{I_NT=0}), since both are simply the current
operator evaluated inside the normal metal. The key physical element
in our `two-fluid' model is that we require that both Eqs.~(\ref{I_C})
and (\ref{2And}) for the `condensate' current $I_C$ and the `Andreev'
current $I_A$ must be equal. We examine this assumption more
rigorously in the Appendix, by evaluating the electrical current
operator inside the superconductor. In this section we make a
plausibility argument for equating $I_A$ from Eq.~(\ref{2And}) with
the `condensate current' $I_C$ from Eq.~(\ref{I_C}), or the `current
of Cooper pairs'.

Eq.~(\ref{2And}) expresses the condensate current in terms of
probabilities for Andreev reflection and Andreev transmisison of an
electron incident from the normal metal. Conversely, Eq.~(\ref{I_C})
expresses the condensate current in terms of the superfluid velocity.
That an incident electron and a reflected hole on the normal metal
side requires a Cooper pair to move off into the superconductor (to
preserve electrical charge conservation) is well
known~\cite{rBTK}. Similarly, Andreev transmission of an electron also
requires a Cooper pair flow inside the superconductor. Andreev
reflection and Andreev transmission of electrons incident from the
normal metal require Cooper pairs flow away from the NS interface (in
the same direction) for both processes.  Equation~(\ref{2And})
embodies this physical reasoning. Similarly, identifying
Eq.~(\ref{I_QP}) as the `quasi-particle' contribution to the
electrical current is also quite natural, being proportional to the
electrical current operator for an electron injected from the normal
metal (evaluated on the superconducting side).

The `two fluid' picture here requires Eqs.~(\ref{I_C}) and
(\ref{2And}) be equal be satisfied in order to guarentee global
current conservation. Once the junction geometry and applied voltage
is specified, the only free parameter in Eqs.~(\ref{I_C}) and
(\ref{2And}) is the superfluid velocity $v_s=\hbar q/m$. Since the
quasi-particle transmission and reflection coefficients themselves
depend on $v_s$, equating Eqs.~(\ref{I_C}) and (\ref{2And}) is then a
globally self-consistent procedure for determining the superfluid flow
velocity $v_s$. Using this value for $v_s$, one then uses either
Eq.~(\ref{I_NT=0}) or (\ref{I_S}) to find the terminal currents for
each value of the bias voltage $V$.

\subsection{Two-Fluid Approximation for Ballistic NS Junction}  
\label{balns}

\indent

\begin{figure}[htb]
\centps{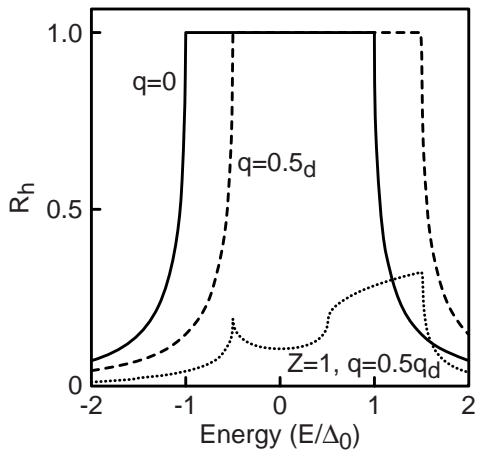}{60}
\caption{Andreev reflection probability for an electron incident on an
NS interface when the superfluid velocity is zero (solid line) and
when $v_S = v_d/2$ (dashed line). Superfluid flow simply shifts the
Andreev reflection probability by an amount $\Delta (v_S/v_{d})$.  If
a tunnel barrier is placed at the NS interface (dotted line), there is
no simple relation between the Andreev reflection probabilities with
and without superfluid flow.}
\label{figAnd}
\end{figure}

We now restrict our attention to ballistic NS junctions, and
approximate the energy bands near the Fermi level as simply rigidly
shifted in energy by an amount $\pm |\Delta| (v_s/v_d)$, as determined
from Eq.~(\ref{shiftE}).  When the superconductor carries a finite
supercurrent, we can then obtain the Andreev reflection coefficient
from
\begin{equation}
T_{Nh,Ne}(E) = R_A^0(E - |\Delta| (v_s/v_d)).
\label{RAshift}
\end{equation}
Here $R_A^0(E)$ is the Andreev reflection coefficient found by
BTK~\cite{rBTK} when the superfluid flow is zero ($v_s=0$), namely
\begin{equation}  
R_A^0(E)= \left\{  
	\matrix{
	1 &  |E| \leq |\Delta| 
	\cr
	\left( 
	\frac{\displaystyle E - \sqrt{E^{2} - |\Delta|^{2}}}
	{\displaystyle E + \sqrt{E^{2} - |\Delta|^{2}}}  
	\right) & |E| \geq |\Delta|
	}
	\right. .
\label{RA0} 
\end{equation}  
This rigid shift in the Andreev reflection coefficient, corresponding
to the rigid shift in the energy bands near the Fermi level, is shown
in Fig.~\ref{figAnd}.  Simply shifting the reflection coefficients in
energy is not a valid approximation when a tunnel barrier is present
at the NS interface~\cite{rRieThes}, as also noted in Fig.~\ref{figAnd}.
Since the differential conductance of an NS junction is 
\begin{equation}
\left. \frac{dI}{dV} \right|_V 
= \frac{2e}{h} [ 1 + R_A (E = eV) - R_N(E = eV) ] ,
\end{equation}
a differential conductance measurement producing a larger than expected
energy gap could point to significant superfluid flow in the junction.

Shifting the Andreev reflection coefficient in energy makes it
possible for the `negative energy' Andreev reflections, i.e. the
Andreev reflection probabilities having $E<0$ in Eqs.~(\ref{RA0}) to
contribute to the excess electrical current. These `negative energy'
Andreev reflections are shown as the additional area under the Andreev
reflection probability for $E>0$ when $v_s > 0$ in Fig.~\ref{figAnd},
At zero temperature, Eqs.~(\ref{I_NT=0}) and (\ref{RAshift}) give
\begin{equation}  
I  = (2e/h) M_N \int^{eV}_{0} 
[1 + R_{A}^0(E - |\Delta| (v_s/v_d))] dE.
\label{INbalT=0}
\end{equation}
To determine the superfluid velocity $v_s$ in Eq.~(\ref{INbalT=0}),
Eqs.~(\ref{I_C}) and (\ref{2And}) require
\begin{equation}  
|\Delta| \alpha (v_s/v_d) 
=  \int^{eV}_{0}  R_{A}^0(E - |\Delta| (v_s/v_d)) dE.
\label{selfconvs}
\end{equation}
Equation~(\ref{selfconvs}) is a self-consistent equation for the
superfluid velocity $v_s$, and depends on the ratio of the number of
conducting modes in the superconductor to the normal conductor $\alpha
= M_S/M_N$. The largest possible excess current would occur if we
could fix $v_s \to \infty$ in Eqs.~(\ref{INbalT=0}), and would give
twice the BTK result of $I_{\rm exc}^{max} = 2 I_{\rm exc}^{BTK} =
(8/3)(2e\Delta/h)$. However, this theoretical maximum excess current is
not possible due to the constraint that the superfluid velocity must
remain smaller than the depairing velocity. One must therefore discard
any solution of Equation~(\ref{selfconvs}) giving $v_s > v_d$.

\subsection{Numerical Evaluation of Two-Fluid Formula}
\label{numeval}

\indent

\begin{figure}[htb]
\twofig{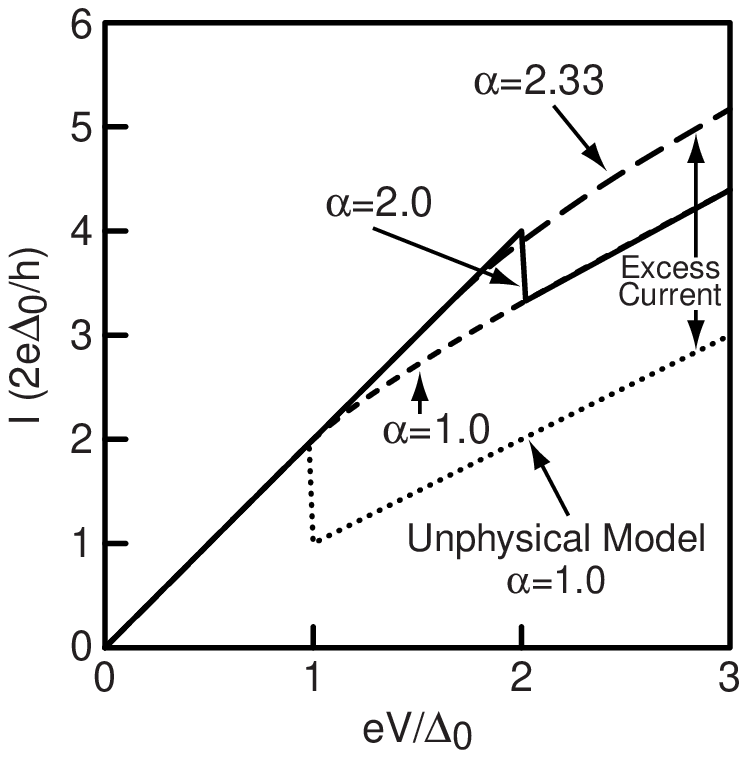}{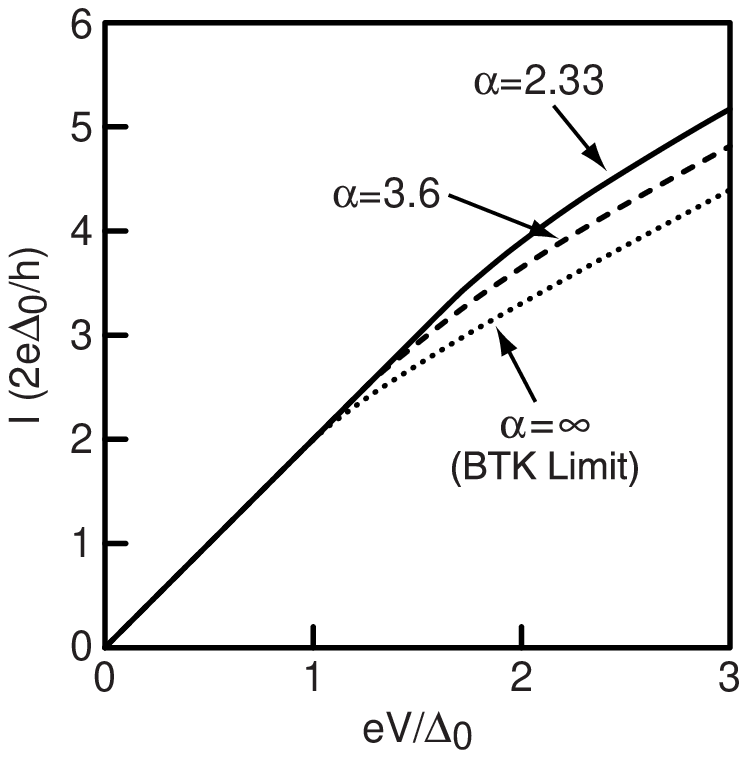}{6.0}
\caption{ (a) When the transverse mode ratio $\alpha < (7/3)$, the
order parameter in the narrow superconducting wire can collapse,
giving rise to discontinuities in the I-V relation. (b) When
$\alpha > (7/3)$ no such discontinuities arise. The excess current
is larger than the BTK result for $\alpha = 7/3$, shown in both
(a) and (b). The I-V evolves smoothly into the BTK result for a
point contact NS junction when $\alpha \to \infty$.}
\label{figIV}  
\end{figure}

We plot the numerical solution of
Eqs.~(\ref{INbalT=0})-(\ref{selfconvs}) for the total current through
the NS junction as a function of the voltage in Fig.~\ref{figIV}.
When the transverse mode ratio is small, namely when $\alpha \leq 7/3$
shown in Fig.~\ref{figIV}(a), there is a voltage above which current
conservation requires that $v_s$ exceed the Landau depairing velocity
of the superconductor. When $v_s \geq v_d$, the narrower
superconducting region (between the large normal contact and the large
superconducting reservoir) becomes a normal conductor.  This collapse
of the order parameter in the narrower superconducting wire continues
until a new NS interface is formed where the narrow conductor meets
the wide superconducting reservoir. Geometrical dilution of the
supercurrent where the superconductor widens into a thermodynamic
reservoir moves the NS interface so that a stable point contact
junction is formed.  Forcing the narrower superconducting region into
the normal state therefore creates a point contact NS junction having
$\alpha = \infty$, i.e. the BTK limit of an NS junction. Forcing the
narrower superconducting region to become normal therefore forces the
excess current to fall abruptly to the BTK limit shown in
Fig.~\ref{figIV}(a).

An excess current larger than the BTK limit is also shown in
Fig.~\ref{figIV}(a), due to the additional `negative energy' Andreev
reflections.  When the transverse mode ratio is larger, namely when
$\alpha \geq 7/3$ shown in Fig.~\ref{figIV}(b), the superfluid
velocity in the narrower superconductor is always less than the Landau
depairing velocity. Consequently, no abrupt drops in the current occur
for any value of the voltage. The excess current simply decreases
gradually from its maximum value at $\alpha = 7/3$ to the BTK value
for a NS point contact at $\alpha = \infty$.

\begin{figure}[htb]
\centps{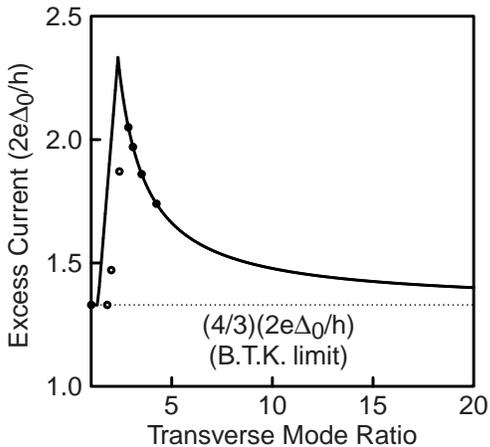}{60}
\caption{The excess current $I_{\rm exc}$ of a ballistic
normal-superconducting wire at zero degrees Kelvin is maximum when the
transverse mode ratio is $\alpha = (M_S/M_N) = 7/3$.  $I_{\rm exc}$
approaches the point contact (BTK) value of $4/3 (2e\Delta/h)$ when
$\alpha \to \infty$.  Collapse of the order parameter in the narrow
superconducting wire limits $I_{\rm exc}$ when $\alpha < 7/3$.}
\label{figExCur}  
\end{figure}

We plot the maximum excess current in a ballistic NS junction versus
the mode ratio $\alpha$ in Figure~\ref{figExCur}. The excess current
at any given voltage is defined as the difference between the current
carried by the NS superconducting junction and a normal NN junction,
namely $I_{ex}(V) = I_{NS}(V) - I_{NN}(V)$.  The maximum excess
current occurs at voltage for which $I_{ex} = {\rm Max} \; [ I_{ex}(V)
]$.  For $\alpha < 4/3$ the excess current is given by the point
contact value, namely the BTK result of $(4/3)(2e\Delta/h)$, and
occurs at a voltage $V = \infty$.  When $4/3 \leq \alpha \leq 7/3$ the
excess current occurs at a finite voltage $V \leq \infty$, and is
larger than the BTK result.  As the transverse mode ratio increased
above $\alpha \geq 7/3$, geometrical dilution of the supercurrent
reduce the band tilting in the superconductor, reducing the maximum
excess current of the junction.  When the mode ratio is very large, so
that $\alpha \to \infty$ we recover the point contact result for the
excess current.

The circled dots in Figure~\ref{figExCur} show the results of a more
realistic numerical calculation~\cite{rRieThes} for the excess
current.  As detailed in Ref.~\cite{rRieThes}, we permit interband
scattering at the NS junction and allow different conducting modes in
the superconductor to carry different amounts of
supercurrent~\cite{rLiFu}. The general behavior for the excess current
versus mode ratio $\alpha$ of this more realistic NS junction model is
quite similar to our simplified model, except the maximum in the
excess current is slightly smaller and shifted to a slightly larger
transverse mode ratio. The slightly lower excess current arises
because the higher lying modes have a smaller Fermi velocity, and
therefore the Andreev reflection coefficients for the higher lying
modes do not shift in energy as much as the lower lying modes.  We
conclude that this more realistic model, even though still a two-fluid
type model which only globally conserves electrical current, confirms
the essential features of our simpler independent mode calculation in
this section.

\subsection{Limiting Cases of Two-Fluid Formula}
\label{limitcases}

\indent

We can understand how the excess current $I_{exc}$ depends on the mode
ratio $\alpha = M_S / M_N$ in Figure~\ref{figExCur} by examining
Eqs.~(\ref{INbalT=0})-(\ref{selfconvs}) in different limits.  The
integral of the shifted Andreev reflection probability in
Eqs.~(\ref{INbalT=0})-(\ref{selfconvs}) can be done analytically to
yield
\begin{equation}  
\int^{eV}_{0}  R_{A}^0(E - |\Delta| (v_s/v_d)) dE
= eV 
\label{intRA01} 
\end{equation}  
when $eV \leq |\Delta| (1 + (v_s/v_d))$. In this limit,
Eqs.~(\ref{selfconvs}) and (\ref{intRA01}) show the superfluid
velocity increases linearly with bias voltage $V$.
For larger biases, namely when $eV \geq |\Delta| (1 + (v_s/v_d))$,
the superfluid velocity increases more slowly with
voltage, as determined from
\begin{eqnarray}
& & \int^{eV}_{0}  R_{A}^0(E - |\Delta| (v_s/v_d)) dE
\nonumber \\
& = & |\Delta| (1 + (v_s/v_d)) + |\Delta| [ \frac{1}{3}
- \frac{1}{2} e^{-\gamma} + \frac{1}{6} e^{-3 \gamma} ] .
\label{intRA02} 
\end{eqnarray}
The factor $\gamma$ in
Eq.~(\ref{intRA02}) is
\begin{equation}  
\gamma = \cosh^{-1}(\frac{eV}{|\Delta|} - \frac{v_s}{v_d}) .
\label{gamma}
\end{equation}  
The excess current we obtain from
\begin{equation}  
I_{exc}  = (2e/h) M_N \int^{eV}_{0} 
R_{A}^0(E - |\Delta| (v_s/v_d)) dE.
\label{IexcT=0}
\end{equation}

Consider first the case where the narrower superconducting wire is not
driven normal, so that $v_s < v_d$ for all values of voltage.  In that
case, the maximum excess current occurs when $V = \infty$, so that
$\gamma = \infty$ in Eq.~(\ref{gamma}).  Eqs.~(\ref{selfconvs}) and
(\ref{intRA02}) for the superfluid velocity then reduce to $(\alpha -
1)(v_s/v_d) = 4/3$. The maximum allowed superfluid velocity,
$(v_s/v_d) = 1$, then occurs for a transverse mode ratio of $\alpha =
7/3$.  The excess current from Eqs.~(\ref{IexcT=0}) and
(\ref{intRA02}) then becomes
\begin{equation}  
I_{exc} = \left( \frac{2e |\Delta|}{h} M_N \right)
\left( \frac{4}{3} \right)
\left( \frac{\alpha}{\alpha-1} \right) ,
\label{IexcT=0high}
\end{equation}
when $\alpha \geq 7/3$. The 
excess current reaches its
maximum value of $I_{exc} = (7/3) (2e |\Delta|/h) M_N$ when
$\alpha \geq 7/3$ as shown in Figure~\ref{figExCur}. Taking the
limit $\alpha \to \infty$ in Eq.~(\ref{IexcT=0high}) recovers the
BTK result $I_{exc} = (4/3) (2e |\Delta|/h) M_N$.

We can also obtain an analytical solution for the excess current using
Eqs.~(\ref{selfconvs}) and (\ref{intRA01}) to determine the superfluid
velocity as $(v_s/v_d) = eV / |\Delta| \alpha$.
The excess current then follows from Eqs.~(\ref{IexcT=0}) and
(\ref{intRA01}) as $I_{exc} = (2e |\Delta|/h) M_N
|\Delta| \alpha (v_s/v_d)$. Using the maximum alllowed depairing velocity
of $(v_s/v_d) = 1$ then gives
\begin{equation}  
I_{exc} = \left( \frac{2e |\Delta|}{h} M_N \right) \alpha.
\label{IexcT=0low}
\end{equation}
The range of allowed mode ratios $\alpha$ for which
Eq.~(\ref{IexcT=0low}) is valid lie along the curve $(v_s/v_d) = 1 =
eV / |\Delta| \alpha$. Furthermore, Eq.~(\ref{intRA01}) is only valid
for $eV / |\Delta| \leq (1 + (v_s/v_d))$. Combining all these
requirements restricts the mode ratio between $1 \leq \alpha \leq 2$.
However, for $1 \leq \alpha \leq 4/3$ the order parameter in the
narrower superconducting wire collapes before the excess current
reaches the BTK value. Eq.~(\ref{IexcT=0low}) therefore describes the
excess current between $4/3 \leq \alpha \leq 2$ as shown in
Figure~\ref{figExCur}.

\section{Exact Solution of a Single Mode NS Junction}
\label{secself1d}

\indent

In this section we wish to evaluate several assumptions made in the
two fluid model of section \ref{secwhymore}. To do this, we solve the
BdG equation, together with the self-consistency requirement for the
order parameter, in a single band (1D) model of the NS junction where
$M_N = M_S = 1$ ($\alpha = 1$).  This self-consistent solution allows
us to demonstrate how the supercurrent flow develops naturally from a
self-consistent solution of the NS junction under a voltage bias. We
use this solution to verify the approximate (two-fluid) procedure we
use to guarentee global current conservation in section
\ref{secwhymore}. One might expect that conserving the current only
globally certainly gives qualitatively correct answers for the I-V
characteristic. However, since the Andreev reflection probability does
not vary much (as a function of energy) if we allow the order
parameter to reach its final self-consistent form, the two-fluid model
also gives accurate quantitative estimates for the I-V
relation. 

Our self-consistent solution of the BdG equations in this section
verifies the main assumptions used in our two-fluid model (when
$\alpha = 1$).  When a voltage is applied to the ballistic NS junction
junction, the magnitude of $\Delta(x)$ remains approximately constant
inside the superconductor (at zero temperature).  However, the order
parameter phase varies approximately linearly inside the
superconducting metal.  There are essentially no additional
quasi-particles injected into the single moded NS junction at zero
temperature, so the total current is simply $I = e n v_s$.  The slope
of the phase is related to the supercurrent velocity as $d \phi/dx =
2q = 2 v_s m / \hbar $. The superfluid velocity is linearly related to
the voltage as $I = (4e^2/h) V = e n v_s$.  As voltage bias increases,
the slope of the phase $d \phi /dx$ increases until the superfluid
velocity reaches the Landau depairing velocity, $v_s = v_d$ (or $d
\phi /dx = 1/\xi_0$). At this voltage $V= V_{c}$ the ordering
parameter inside the narros superconductor collapses, and a stable
point contact NS junction forms inside the wide superconducting
reservoir.

\subsection{Self-Consistent Solution Procedure for BdG Equation}
\label{selfconproc}

The motion of quasi-particles in our one band NS junction, including
the superfluid flow inside in the superconductor, is determined from
the 1D time independent Bogliobov-de Gennes~\cite{rBdG} (BdG)
equation
\begin{equation}  
\left(  
	\matrix{
        H(x)-\mu & \Delta(x) \cr
        \Delta^*(x) & -(H^*(x) - \mu )  \cr}
\right) 
\left(  
	\matrix{
        u(x) \cr   
        v(x) }
\right)  
= E 
\left(  
	\matrix{
        u(x) \cr   
        v(x)}
\right)  
\label{eBdG}  
\end{equation}  
The one-electron Hamiltonian $H(x)$ in Eq.~(\ref{eBdG}) is  
\begin{equation}  
H(x) = -\frac{\hbar^2}{2m} \frac{d^{2}}{dx^{2}}  +  V(x)  .
\end{equation}  
The order parameter $\Delta(x)$ in Eq.~(\ref{eBdG}) is given by
\begin{eqnarray}
\Delta(x) & = & g(x) F(x) \nonumber \\
& = & g(x) \sum_{pn} v^{*}_{pn}(x) u_{pn}(x) f_{pn}
\theta(|E_{pn}| - \hbar \omega_D),
\label{eCond}
\end{eqnarray}
where $g(x)$ is the electron-phonon interaction strength at each point
and $F(x)$ is the pair correlation function. (Although we assume $g(x)$
is local, in reality it is spread over a correlation 
distance $v_F/\omega_D$.) The index $p$ in
Eq.~(\ref{eCond}) denotes the lead from which the scattering state
originates, namely $p=N$ is the left lead and $p=S$ the right lead.
The index $n$ in Eq.~(\ref{eCond}) denotes the good quantum numbers in
the lead, namely $n = (k,\beta)$ where $k$ is the wavenumber and
$\beta = (e,h)$ the electron-like and hole-like states.  The sum in
Eq.~(\ref{eCond}) runs over states injected from the leads, including
both positive ($E_n > 0$) and negative energies ($E_n < 0$).  The
coherence factors $u(x)$ and $v(x)$ in Eq.~(\ref{eCond}) are functions
of the order parameter $\Delta(x)$ through Eq.~(\ref{eBdG}).

In this section we show the self-consistent solutions of
Eq.~(\ref{eBdG}) and (\ref{eCond}) for a voltage-biased NS
junction. Details of the self-consistent solution procedure are given
in Ref.~\cite{rRie1}.  To solve the order parameter self-consistently,
we first assure an initial or zeroth order guess $\Delta_0(x)$ for the
order parameter.  We then divide the one dimensional space is into
differential elements, where the magnitude of the order parameter
superfluid velocity are constant in each section. We match the
wavefunctions and their derivatives at each interface to obtain the
zeroth order wavefunctions $u_0(x)$ and $v_0(x)$. The first iteration
for the order parameter $\Delta_1(x)$ we then obtain from
Eq.~(\ref{eCond}) using the zeroth order wavefunctions $u_0(x)$ and
$v_0(x)$, etc. The zeroth order guess for the order parameter
$\Delta_0(x)$ can either be constant, i.e. $\Delta_0(x) = \Delta$, or
it can contain a superfluid flow, i.e. $\Delta_0(x) = \Delta
e^{2iqx}$. Given the same electrical current flow, either initial
guess for the order parameter converges to the same final answer.

The voltage bias across the NS junction is a boundary condition which
determines the Fermi occupation probabilities $f_{pn}$ in
Eq.~(\ref{eCond}). The occupation probability $f_{pn}$
of a scattering state $(p,n)$ which originates 
inside the normal or superconducting reservoir $p$, is the same for
holes and electrons when the applied bias is zero ($V=0$).  Under a
voltage bias, however, electrons in the normal metal are occupied up
to an energy $\mu+eV$, while holes are occupied up to an energy
$\mu-eV$.~\cite{rDatta1} These different Fermi factors electron-like
and hole-like quasi-particles injected from the normal metal are shown
schematically in Fig.~\ref{figBand}. The unequal occupation
probabilities for holes and electrons injected from the normal metal
causes these two classes of scattering states to contribute
differently to the sum in Eq.~(\ref{eCond}) under an applied bias.
We can write the Fermi factors as
\begin{equation}
f_{pn} = f_{p \beta} = f(E - eV_{p \beta} ) \; ,
\label{fermi}
\end{equation}
where $f(E) = 1/[1+{\rm exp} \left( E / k_B T \right)]$. Here
$eV_{p \beta}$ is
effective biasing voltage (or effective electrochemical potential)
applied to the $(p \beta)$th
lead, namely~\cite{rDatta1}
\begin{equation}
V_{Ne} =  V \; ,
\label{vefactors}
\end{equation}
\begin{equation}
V_{Nh} =  - V \; .
\label{vhfactors}
\end{equation}
In this paper the superconducting leads are grounded so that $V_{S
\beta} = 0$.

To obtain the electrical current $I(x)$, we do not invoke any ad hoc
`source term' as done in Ref.~\cite{rBTK}, but instead simply evaluate
the electrical current operator for a scattering state originating in
lead $q$ (having quantum number $n$) and terminating in lead $p$,
namely
\begin{equation}
I_{q} =  \sum_{pn} 
\left( J_{u} + J_{v} \right)_{q ; pn} f_{pn}
-  \sum_{p n} \left( J_{v} \right)_{q ; p n}
\; .
\label{eCur}
\end{equation}
The $J_{u}$ and $J_{v}$ are Schr\"{o}dinger currents associated with
the waves $u$ and $v$, namely $J_{u}=(e\hbar/m) {\rm Im}\{u^{*}(x)
\nabla u(x)\}$ and $J_{v}=(e\hbar/m) {\rm Im}\{v^{*}(x) \nabla
v(x)\}$.  The `vacuum current' due to the filled hole band is argued
in Ref.~\cite{rDatta1} to be zero, namely $\sum_{q n} (J_{v})_{p ; q
n} = 0$, as we have confirmed for the NS junction. Solving
Eq.~(\ref{eBdG}) together with Eq.~(\ref{eCond}) guarentees electrical
current
conservation~\cite{rKum,rMathews},\cite{rBag1},\cite{rSols,rFurusaki3},
even when the superconductor is far from equilibrium. A proof of this
statement for NS junctions follows from generalizing the discussion in
Appendix B of Ref.~\cite{rBag1} to the nonequilibrium case. If the
nonequilibrium system involves two superconductors at different
biases~\cite{rHurd}, current conservation is more complex and $\sum_{q
n} (J_{v})_{p ; q n} \ne 0$.

\subsection{Order Parameter Phase}
\label{opphase}

Consider first the NS junction, where the coupling constant $g(x)$ is
\begin{equation}
g(x) = \left\{
	\matrix{
	g_R & x > 0 \cr
	0 & x < 0 }
     \right.
\end{equation}
We choose $g_R$ and $\omega_D$ so that the critical temperature of the
right superconductor is $T_c = 6.6$K. Our initial guess for the
order parameter we take to be
\begin{equation}
\Delta_0(x) = \left\{
	\matrix{
	\Delta & x > 0 \cr
	0       &   x < 0 }
     \right.
\label{delta0}
\end{equation}
We therefore do not force a superfluid flow inside the superconductor
from our zeroth order guess for the order parameter.

\begin{figure}[htb]
\twofig{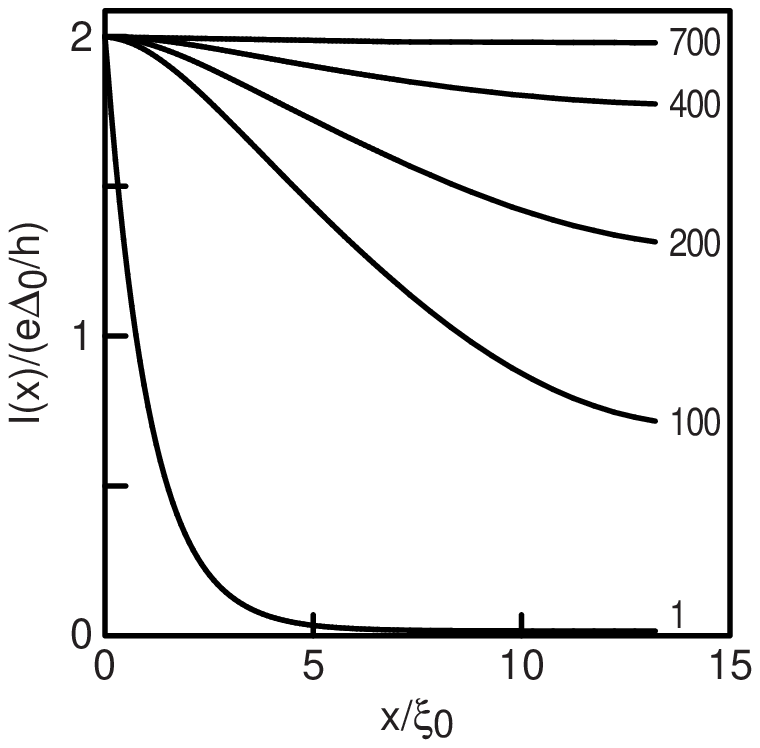}{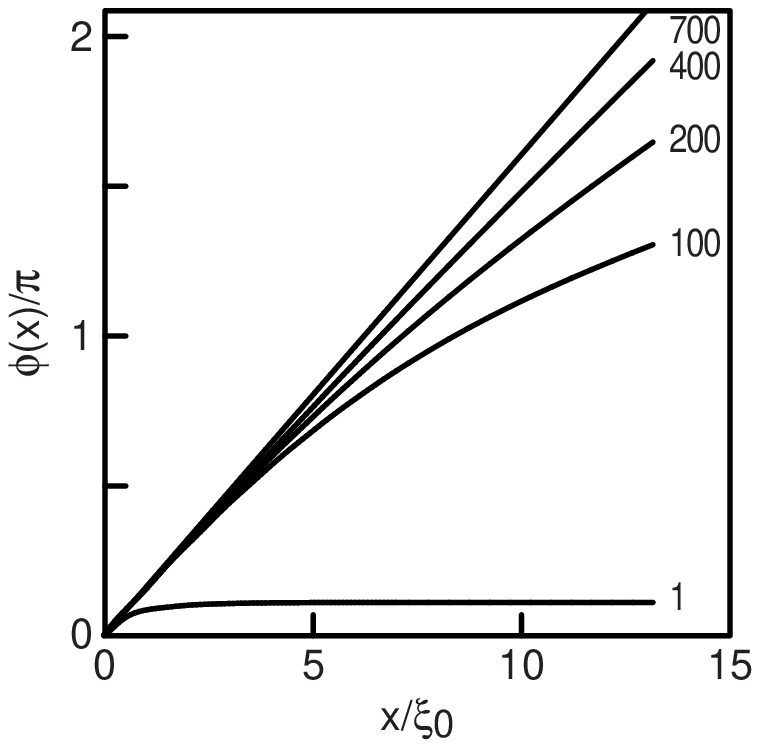}{6.0}
\caption{Both (a) the electrical current throughout an NS junction,
and (b) the phase in the superconducting metal, develop naturally from
a self-consistent model. The initial order parameter guess
$\Delta_0(x)$ assumed zero supercurrent, however a superfluid flow
appeared naturally upon reaching self-consistency.}
\label{figcurrcons}
\end{figure}

Fig.~\ref{figcurrcons}(a) shows the electrical current $I(x)$ versus
position $x$ inside the superconductor. The numbers beside the lines
in Fig.~\ref{figcurrcons}(a) denote the iteration number. For the
first iteration ($N=1$), the electrical current dies off within a
coherence length of the NS interface, so that the electrical current
is not conserved. After the iterative scheme converges ($N=700$) in
Fig.~\ref{figcurrcons}(a), we see the electrical current $I(x)$ is
constant as a function of position, indicating the electrical current
is indeed conserved. In Fig.~\ref{figcurrcons}(b) we plot the order
parameter phase inside the superconductor as a function of position.
A uniform phase gradient develops inside the superconductor when the
iterative scheme has converged to self-consistency, showing that the
development of a supercurrent is necessary to guarentee current
conservation. We expected this constant order parameter phase gradent
in the NS junction, since it is similar to that constant order
parameter phase found in the self-consistent solution of the SNS
junction.~\cite{rRie1}

We can understand these result using our two-fluid picture from
section \ref{secwhymore}, and assuming rigidly shifted energy
bands. The maximum current of $4e\Delta/h$ is reached when the voltage
$eV = \Delta$. At this point the energy bands have shifted up faster
than the Fermi level of the normal contact, and thus no direct
transmission of electrons across the junction inside the energy range
$0<E<eV$ is allowed. Andreev transmission will also be small, as is
usually the case in NS junctions. There will therefore be essentially
no quasi-particles above the Fermi level of the superconductor, while
all states below the Fermi level are filled. The quasi-particle
contribution to the current inside the superconductor, $I_{QP}$ in
Eq.~(\ref{I_QP}), is essentially zero in the single mode NS junction
when $eV < \Delta$. The electrical current is therefore $I = e n v_s$
in this single moded NS junction, the same as for a uniform 1D
superconductor. The order parameter magnitude collapses when the
superfluid velocity equals the depairing velocity $v_s = v_d$, at a
bias voltage $eV = \Delta$.

\subsection{Order Parameter Magnitude}
\label{opmag}

\indent

\begin{figure}[htb]
\twofig{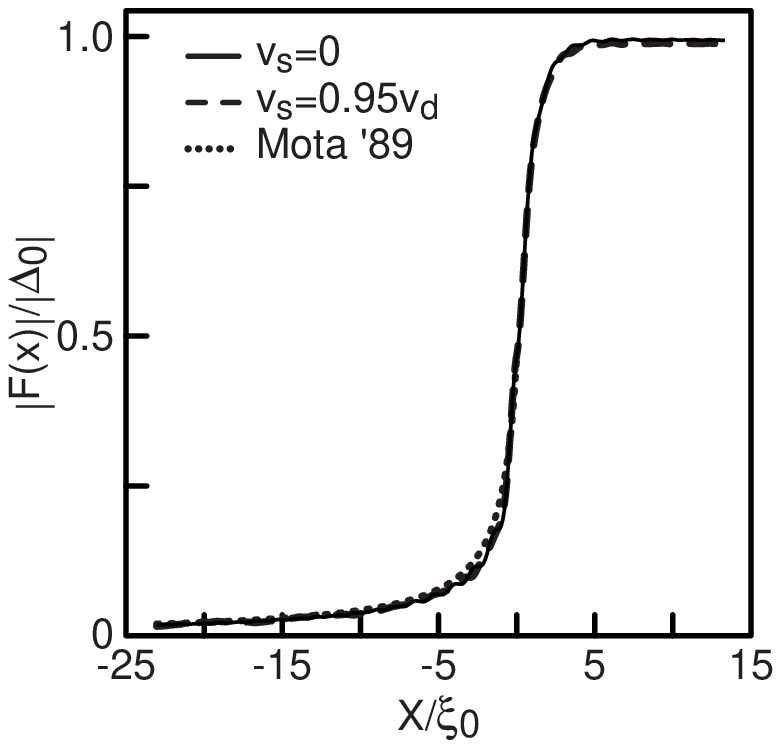}{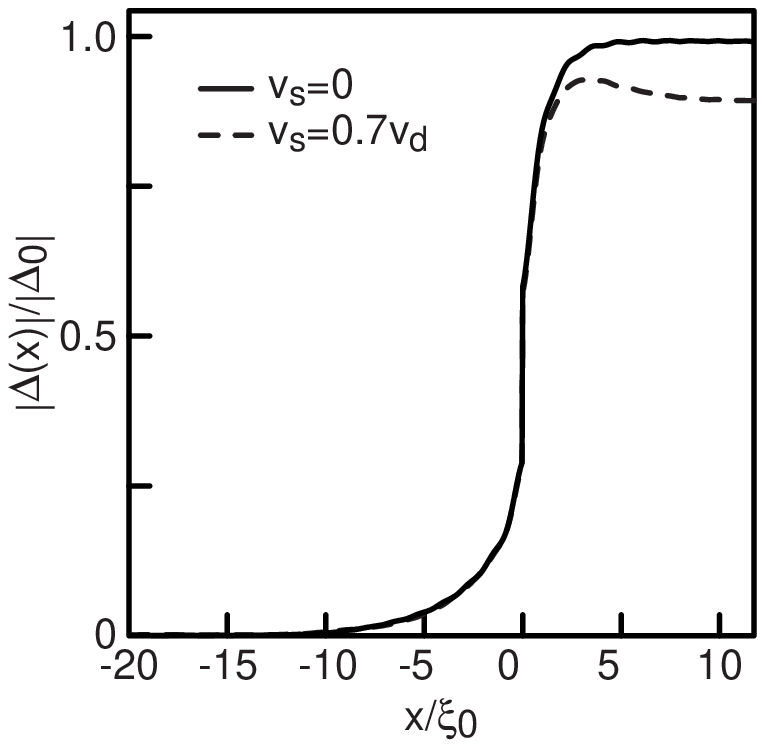}{6.0}
\caption{The magnitude of the coherence function $F(x)$ (a) changes
little when a large flow is present in a NS junction with an applied
voltage.  The solid line is $F(x)$ when the applied voltage is zero.
When the applied voltage is large $.95\Delta$ $F(x)$ shows little
change (dashed line) from the junction in equilibrium.  [The phase of
$F(x)$ changes linearly with postion throughout the NS junction.]  (b)
In an S'S junction at a temperature above the critical temperature of
S', the order parameter is non-zero in the "normal" metal.  The finite
temperature in combination with a moderate supercurrent causes a
supression of the larger gap superconductor.  (dashed line)}
\label{figorderparam}
\end{figure}

Fig.~\ref{figorderparam}(a) shows the magnitude of the condensation
amplitude $F(x)$ as a function of position in both the normal and
superconducting metal at zero temperature.  The solid line indicates a
bias voltage of $V=0$, while the dashed line is for a bias voltage $V
= 0.95 V_c$ ($v_s = 0.95 v_d$). The general form of the condensation
amplitude $F(x)$ for a ballistic NS junction at equilibrium is well
known from earlier non-self-consistent
models.~\cite{rMacMill},\cite{rFalk}.  We find substantial agreement
between these earlier results and our fully self-consistent
calculations. In the superconductor, $F(x)$ is suppressed from its
bulk value near the NS interface. In the normal metal, $F(x)$ shows
behavior quite similar to the low temperature experimental results of
Mota~\cite{rMota}.  The dotted line in Fig.~\ref{figorderparam}(a) is
$F(x) \propto 1/(|x| + x_{0})$, found experimentally by
Mota~\cite{rMota}.  The value of $x_{0}$ used in
Fig.~\ref{figorderparam}(a) is $x_{0} = \xi_{0}$. The fit between the
experimental determined form $F(x) \propto 1/(|x| + x_{0})$ and the
results of our self-consistent calculation is quite good.  The result
$F(x) \propto 1/(|x| + x_{0})$ for the pair correlation function at
low temperature in an NS junction at equilibrium ($V=0$) was also
pointed out by Falk~\cite{rFalk} for the asymptotic limit $x
\rightarrow \infty$.

In an NS junction, the ordering parameter $\Delta(x)$ vanishes in the
normal metal because the electron-phonon coupling constant $g(x)$ is
zero there.  For a finite ordering parameter $\Delta(x)$ to exist
inside the normal metal we must have $g(x) \ne 0 $ in the normal
metal. One way to achieve a non-zero $g(x)$ in the normal metal is to
fabricate an $S^{\prime}S$ junction, where the superconductor
$S^{\prime}$ has a smaller critical temperature than S. If we then
elevate the temperature so that $T_c > T > T_c^{\prime}$, we
effectively form an NS junction where $g(x)$ is not zero inside the
normal metal.  Fig.~\ref{figorderparam}(b) shows a self-consistent
calculation for an such $S^{\prime}S$ junction, where $T_c = 6.6K$,
$T_c^{\prime}= 0.66K$, and $T=2K$. Unlike the NS junction, where
$\Delta(x)=0$ inside the normal metal, we see a non-zero `tail' of the
ordering parameter extending into the normal metal in
Fig.~\ref{figorderparam}(b). At zero temperature, the bulk value of
the order parameter inside the weaker superconductor is
$\Delta^{\prime}(T=0) = 0.1 \Delta (T=0)$. From
Fig.~\ref{figorderparam}(b) we see that $\Delta(x=0, T=2K) \simeq 2.5
\Delta^{\prime}(T=0)$, larger than even the bulk value of the order
parameter in the weaker superconductor at zero temperature. When a
voltage is applied to the $S^{\prime}S$ junction, namely
$eV=.7\Delta_{0}$ in Fig.~\ref{figorderparam}(b), the ordering
parameter inside $S$ is now suppressed from its bulk value at
$T=2K$. This degradation of the order parameter at finite temperature,
when the superconductor carries a finite supercurrent, is similar to
that of a bulk superconducting wire~\cite{rBag1}. The tail of the
ordering parameter extending into $S^{\prime}$ is only slightly
changed in the presence of the supercurrent.

\subsection{Local Density of States}
\label{locdos}

\indent

\begin{figure}[htb]
\twofig{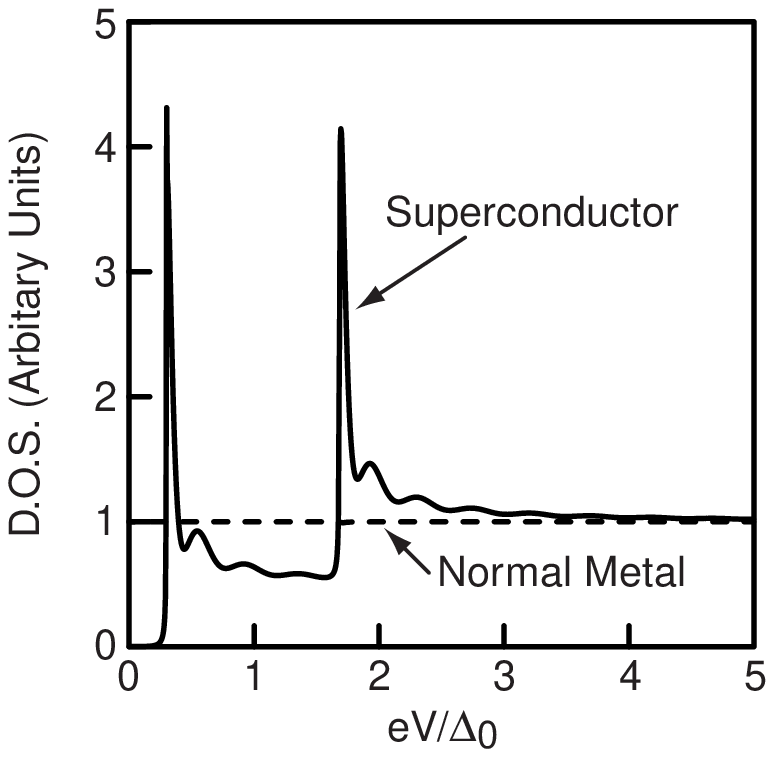}{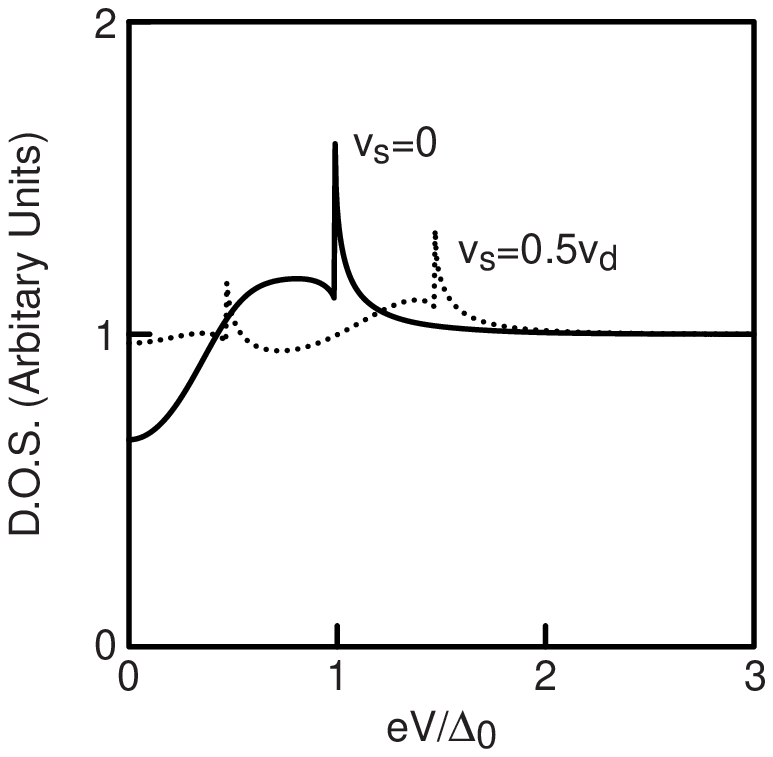}{6.0}
\caption{(a) When a supercurrent is present in an NS junction, the
local density of states inside the superconductor shows two peaks. The
local density of states inside the normal conductor is constant for
all $x < 0$, indicating the lack of any energy gap due to the
proximity effect. (b) In an S'S junction when a
supercurrent is present, where S' is a superconductor above its 
transition temperature, structure is seen in the local density of states
inside the "normal" S' metal. }
\label{figDOS}
\end{figure}

In addition to the magnetic susceptibility techniques used by Mota,
which explore the condensation amplitude $F(x)$ in the normal metal,
another method to experimentally investigating how the ordering
parameter $\Delta(x)$ varies near NS interfaces is tunnelling
spectroscopy using an STM tip.~\cite{rTess}. We expect that a
measurement of the differential conductance $dI/dV$ at the STM tip is
proportional to to the local density of states $N(x,E)$.  We can
calculate the local density of states 
using the equation
\begin{equation}
N(x,E) = \frac{1}{\pi}\sum_{p,n}
{|u_{p,n}(x,E)|^{2} + |v_{p,n}(x,E)|^{2}}|\frac{dk}{dE}| ,
\end{equation}
where $p$ is the lead index and the quantum number $n = (k,\beta)$.
Fig.~\ref{figDOS}(a) shows the local density of states for an NS junction
having an applied voltage of $eV=.7\Delta_{0}$.  At this bias voltage,
the superconductor carries a supercurrent of $v_s = 0.7 v_{d}$.  The
solid line shows $N(x,E)$ at a position $x = 5 \xi_0$ inside the
superconductor. The original peak in the density of states, which
occurs at $E = \Delta$ when the superfluid flow $v_s = 0$, splits into
two separate peaks. As the energy bands inside the superconductor tilt
under the superfluid flow, the band edges move to the energies $E= [1
\pm (v_s/v_{d})]\Delta$, as do the peaks in the density of states.  In
the normal metal, the density of states is approximately constant for
energies of interest.  The constant density of states in the normal
metal is due to a zero pairing potential $\Delta(x) = 0$ inside the
normal metal, since $g(x)=0$ in the normal metal.

Fig.~\ref{figDOS}(b) shows the local density of states for an
$S^{\prime}S$ junction having $T_c > T > T_c^{\prime}$. We evaluate
the local density of states at a position $x = -\xi_{0}$ in the normal
metal. The local density of states inside the stronger superconductor
is approximately the same as Fig.~\ref{figDOS}(a).  For the two
different applied voltages, where $v_{s}=0$ (solid) and
$v_{s}=.5v_{d}$ (dotted), the presence of the superconductor changes
the density of states inside the normal metal. For $v_{s}=0$, there is
a depression in the density of states near $E=0$, showing the partial
development of an energy gap inside the normal metal. The density of
states does not go to zero at $E=0$, since quasi-particles incident
from the left contact can still propagate to the position $x =
-\xi_{0}$.  As the current increases, the density of states inside the
normal metal becomes flatter due to injection of quasi-particles from
the tilted energy bands inside the superconductor.  The two small
peaks at $eV=.5\Delta_{0}$ and $eV=1.5\Delta_{0}$ when $v_{s}=.5v_{d}$
are again associated with the tilted energy bands inside S, and are
significantly broadened by thermal smearing at $T= 2K$ (which we have
ignored in Fig.~\ref{figDOS}(b)).

To summarize, while looking to justify our two-fluid model for the
effect of superfluid flow in an NS junction, we studied the pair
corrlation function $F(x)$ and ordering parameter $\Delta(x)$.
We looked at $F(x)$ and
$\Delta(x)$ inside both (1) an NS junction at temperature $T=0$, and
(2) an S$^{\prime}$S junction at a temperature $T_c > T >
T_c^{\prime}$. Here S$^{\prime}$ is a superconductor having an order
parameter smaller than $S$, so that $T_c^{\prime} < T_c$. The
S$^{\prime}$S junction at this temperature is therefore a type of NS
junction. First, the condensation amplitude is $F(x) \simeq x_0 / (|x|
+ x_0)$, approximately independent of the applied voltage.  Second, at
the voltage $V_{c} = \Delta_{0}/e$, the ordering parameter of the
superconductor collapses, i.e. $\Delta_0 \to 0$. The supercurrent
carried inside the superconductor at a voltage $V = V_{c}$ is
approximately equal to the Landau depairing current $I_C = e n
v_d$. Third, we show how the supercurrent changes the 1-D local
density of states per unit energy $N(x,E)$ at various points in the
normal and superconducting metals, both for the NS junction and the
S$^{\prime}$S junction. The local density of states shows the
influence of the superfluid flow.

\subsection{Locally (But Not Globally) Gapless Superconductivity}
\label{gapless}

The phase gradient of $F(x)$ is approximately constant for both the NS
junction in Fig.~\ref{figorderparam}(a) and the $S^{\prime}S$ junction
in Fig.~\ref{figorderparam}(b). Since the energy gap $\Delta(x)$
decays to zero in Fig.~\ref{figorderparam}(a)-(b), there are regions
of local `gapless' superconductivity near the NS interface where $v_s
> \Delta(x) / p_F$. However, these regions of local `gapless'
superconductivity do not affect bulk properties such as critical
current, critical temperature, etc. Since our self-consistent model
shows that the supercurrent is approximately constant throughout the
superconductor, depairing will occur when $v_s=v_{d}$ everywhere in
the superconducting wire.

A different type of `global' gapless superconductivity has been
discussed by Sanchez and Sols~\cite{rSanSol}. In this proposed gapless
superconductor~\cite{rSanSol}, a superconducting wire in contact with
an NS junction is postulated to exist when $v_{s} > v_{d}$. It is true
that such a non-equilibrium self-consistent solution to the BdG
equations with $v_{s} > v_{d}$ does exist for a uniform
superconducting wire. However, connecting such a wire in this novel
`global' gapless superconducting state to an NS junction imposes the
additional constraint that the self-consistency condition in
Eq.~(\ref{eCond}) must be satisfied at every point in space. When
Eq.~(\ref{eCond}) is satisfied, the magnitude of the order parameter
can no longer be a constant in space, as required for the globally
gapless solution proposed in Ref.~\cite{rSanSol} to exist. 

Another way to view the situation proposed in Ref.~\cite{rSanSol} is
that, for a given applied voltage, the constraint of current
conservation fixes both $\Delta(x)$ and $v_s$, leaving no more degrees
of freedom.  The order parameter $\Delta(x)$ and superfluid velocity
$v_s$ cannot be adjusted independently as required for the bulk
solution of Ref.~\cite{rSanSol} to exist in an NS junction.  Our
self-consistent model shows instead that the ordering parameter of the
superconducting wire collapses when the junction voltage is
approximately $eV = \Delta$, or equivalently when $v_s = v_d$.  In
short, we see no possible way to achieve the non-equlibrium conditions
necessary for this novel gapless superconducting state of
Ref.~\cite{rSanSol} to exist by connecting a superconducting wire to a
ballistic NS junction. Fortunately, many other interesting
measurements are possible at ballistic NS interfaces without invoking
the global gapless superconducting state of Ref.~\cite{rSanSol}. In
particular, the excess current larger than the point contact limit,
which we find in section \ref{secwhymore}, in no way depends on the
gapless state proposed in Ref.~\cite{rSanSol}.

\section{Conclusions}  
\label{conc}

\indent

It is possible to experimentally observe excess currents larger than
the $(4/3)(2e \Delta/h)$ found for ballistic NS point contacts. By
varying the width $W_S$ of a superconducting wire in contact with a
normal metal having width $W_N$, one can vary the effect of the
supercurrent on the energy bands in the superconductor.  Varying the
widths of the two conductors controls the ratio of the number of
conducting modes $\alpha \simeq W_S / W_N$ in the ballistic NS
junction. We find the excess current attains a theoretical maximum of
$(7/3)(2e \Delta/h)$ when $\alpha \simeq 7/3$. For $1< \alpha < 7/3$
it should be possible to observe discontinuities in the I-V relation
of the NS junction when the superfluid velocity $v_{s}$ exceeds the
Landau depairing velocity $v_d$. Although these results follow from a
simple model which treats all conducting modes as equivalent, we
confirmed the qualitative results using a more realistic model which
includes the different supercurrent carried in each conducting mode
and the scattering between the different modes.  These predictions are
based on a `two-fluid' solution of the BdG equations, in which current
conservation is violated locally near the NS interface.

To confirm our that our `two-fluid' type treatment of the superfluid
flow in NS junctions generates qualitatively accurate predictions for
the I-V relations, we solved the BdG equations self-consistently for a
single mode NS junction under an applied bias. Current conservation
follows automatically in this self-consistent scheme, and shows that
the superfluid velocity is indeed constant throughout the NS junction.
The two important features confirmed in this self-consistent solution
are (1) the superfluid velocity and terminal currents are the same as
required by our `two-fluid' scheme and (2) the order parameter indeed
collapses when the superfluid velocity $v_{s}$ equals the depairing
velocity $v_d$. We did not perform a completely self-consistent
calculation for the multiple moded NS junction, as we believe all the
essential elements of this problem (so far as the excess current is
concerned) are encompassed in the two-fluid treatment of the
multiple-moded NS junction.

Having obtained the a self-consistent solution of the BdG equations
for an NS junction under bias also enabled us to study the pair
correlation function $F(x)$, order parameter $\Delta(x)$, and local
density of states $N(x,E)$ in the NS junction. The peak near the
superconducting energy gap at $E = \Delta$ in local density of states
$N(x,E)$ inside the superconductor is split by the superfluid flow, as
can be measured using STM spectroscopy. If the electron-phonon
interaction in the normal metal is nearly zero, the density of states
in the normal metal is unaffected by the presence of the
superconductor.  If the normal metal $N$ is a weak superconductor held
above its transition temperature, the STM can also measure changes in
the local density of states $N(x,E)$ inside the normal metal. Due to
the uniform superfluid velocity inside an NS junction, and the
reduction of the order parameter near the NS interface, the Landau
depairing condition is locally violated and a type of gapless
superconductivity occurs locally near the NS interface.

\section{Acknowledgments}
\indent

We thank Supriyo Datta for many useful discussions.  We gratefully
acknowledge support from the David and Lucile Packard Foundation and
from the MRSEC of the National Science Foundation under grant
No. DMR-9400415 (PFB).

\section{Appendix: Determining the Superfluid Flow Velocity}
\label{AppDetFlow}

If the pairing potential and electron wavefunctions are determined
from a self-consitent solution of the BdG equations, (1) electrical
current will be conserved everywhere in space and (2) the superfluid
flow will develop naturally as the scheme evolves towards
self-consistency (c.f. section \ref{secself1d}).  In this appendix we
examine a different `globally self-consistent' scheme, where current
conservation is only guarenteed at the device leads. In this scheme,
the correct value of the superfluid velocity $v_S = \hbar q /m$ is
determined by equating the current operator evaluated on the normal
side of the NS junction with the same operator evaluated deep (several
coherence lengths) inside the superconductor. We then adjust the
superfluid velocity (which is a free parameter in the scheme) until
the current flowing out of the superconducting lead is the same as the
current flow in the normal lead. We consider only a single moded 
NS junction with $M_N = M_S = 1$, although the results here are easily
generalized to multiple conducting modes.

The derivation of the electrical current in Ref.~\cite{rDatta1}
applies for an NS junction subject to a superfluid flow. Evaluating
the current operator inside the normal region, Ref.~\cite{rDatta1}
finds the well known formula developed by BTK~\cite{rBTK}, namely
\begin{eqnarray}
I = (2e/h) \int^{\infty}_{-\infty} & &
[1 - T_{Ne,Ne}(E)  + T_{Nh,Ne}(E)] \nonumber \\
& & [f(E-eV) - f(E)] dE.
\label{I_N}
\end{eqnarray}
Here we use the notation of Ref.~\cite{rDatta1}, where $T_{Ne,Ne}=R_N$
and $T_{Nh,Ne}=R_A$ are the particle current reflection probabilities
for an electron-like quasi-particle incident from the normal metal
(right index, `$Ne$') to transmit as a hole or electron in the normal
metal (left index). The sum rule from Eq.~(\ref{Pconserv}) then gives
$I = I_{QP} + I_A$, with
\begin{eqnarray}
I_{QP} = (2e/h) \int^{\infty}_{-\infty} & & 
[T_{Se,Ne}(E) - T_{Sh,Ne}(E)] \nonumber \\
& & [f(E-eV) - f(E)] dE,
\label{I_QPTne0}
\end{eqnarray}
and
\begin{eqnarray}
I_A = (2e/h) \int^{\infty}_{-\infty} & &
[T_{Nh,Ne}(E)+ T_{Sh,Ne}(E) ] \nonumber \\
& & [f(E-eV) - f(E)] dE .
\label{2AndTne0}
\end{eqnarray}

To evaluate the electrical current operator inside the superconductor,
we first note that the energy bands inside superconductor subject to a
superfluid flow are~\cite{rBag1}
\begin{eqnarray}  
\frac{\hbar^{2}k^{2}}{2m} + \frac{\hbar^{2}q^{2}}{2m} - \mu 
= \pm \sqrt{(E \mp \hbar^{2} kq/m)^{2} -  |\Delta|^{2}} \; .
\label{dispersion}
\end{eqnarray}  
The discussion in Ref.~\cite{rBag1} can be extended to show that the
particle current incident from the superconductor $J_P = (1/\hbar)
(dE/dk)$. Thus, the particle current incident from the superconductor
per unit energy is simply $1/h$.  Quasi-particles incident from the
superconductor will then carry an electrical current of the form
\begin{eqnarray} 
I = \frac{e}{h} \int dE \frac{J_P^{\rm out}}{J_P^{\rm in}}
\frac{J_Q^{\rm out}}{J_P^{\rm out}} .
\label{exampleJQ}
\end{eqnarray}
We recognize the particle current transmission coefficient in
Eq.~(\ref{exampleJQ}) as 
\begin{equation}
T_{\rm out, in} = \frac{J_P^{\rm out}}{J_P^{\rm in}}.
\label{defTP}
\end{equation}
To evaluate the $J_Q^{\rm out}/J_P^{\rm out}$ term in
Eq.~(\ref{exampleJQ}) we use several results from Ref.~\cite{rBag1}.
The scattering states inside the superconductor have solutions of the
form
\begin{eqnarray}  
\left(  
	\matrix{
        u(x) \cr  
        v(x) }
\right)  
=
\left(  
	\matrix{
        u_{kq} \exp (iqx) \cr
        v_{kq} \exp (-iqx) }
\right)  
e^{ikx}.
\label{uvkq}
\end{eqnarray}  

{ \onecolumn

The electrical current carried by each occupied state is therefore
\begin{equation}
J_Q = \left( \frac{\hbar k}{m} \right)
\left( |u_{kq}|^2 + |v_{kq}|^2 \right) f 
+ \left( \frac{\hbar q}{m} \right)
\left( |u_{kq}|^2 - |v_{kq}|^2 \right) f ,
\label{defJQ}
\end{equation}
and the particle current for each occupied state is
\begin{equation}
J_P = \left( \frac{\hbar k}{m} \right)
\left( |u_{kq}|^2 - |v_{kq}|^2 \right) f
+ \left( \frac{\hbar q}{m} \right)
\left( |u_{kq}|^2 + |v_{kq}|^2 \right) f .
\label{defJP}
\end{equation}
The state are normalized so that $|u_{kq}|^2 + |v_{kq}|^2 = 1$.
Hence the factor
\begin{equation}
\frac{J_Q^{\rm out}}{J_P^{\rm out}} =
\frac{1}{(|u_{kq}|^2 - |v_{kq}|^2) + (q/k)}
+ \frac{(q/k)(|u_{kq}|^2 - |v_{kq}|^2)}
{(|u_{kq}|^2 - |v_{kq}|^2) + (q/k)}.
\label{defJQ/JP}
\end{equation}
The first term in Eq.~(\ref{defJQ/JP}) one can show is simply the
ratio of the density of states in the superconductor to that of the normal
metal (for a fixed value of wavevector $k$, not a fixed energy), namely
\begin{equation}
\tilde{N}^S(E) \equiv \frac{N_S(k)}{N_N(k)} = 
\frac{1}{(|u_{kq}|^2 - |v_{kq}|^2) + (q/k)} .
\label{dosk}
\end{equation}
We make the translation between wavevector $k$ and energy $E$ inside
the superconductor using Eq.~(\ref{dispersion}).  The second term in
Eq.~(\ref{defJQ/JP}) will be only a minor correction to the first,
being nearly zero near the Fermi level, equal to $(q/k) << 1$ over
most of the energy range, and equal to 1 only near the bottom of the
electron energy bands. The first term in Eq.~(\ref{defJQ/JP}) therefore
dominates, being much larger than 1 near the Fermi level and equal to
1 over most of the energy range.

Applying the procedure outlined in Ref.~\cite{rDatta1} for
construction of the scattering states, and multiplying by their
appropriate Fermi occupation factors, we find a total current inside
the superconductor of
\begin{equation}
I_S = I_1 + I_2 .
\label{IS}
\end{equation}
The current $I_1$ arises from first term in Eq.~(\ref{defJQ/JP})
and is
\begin{eqnarray}
I_1 & = & \frac{2e}{h} \int_{-\infty}^{\infty} 
[\tilde{N}^S_{Se,out}(E) T_{Se,Ne}(E)
- \tilde{N}^S_{Sh,out}(E) T_{Sh,Ne}(E)] 
[f(E-eV) - f(E)] dE
\nonumber \\
& + & \frac{2e}{h} \int_{-\infty}^{\infty} \tilde{N}^S_{Se,out}(E) [2f(E)-1] dE
- \frac{2e}{h} \int_{-\infty}^{\infty} \tilde{N}^S_{Se,in}(E) [2f(E)-1] dE .
\label{I1s}
\end{eqnarray}
To obtain Eq.~(\ref{I1s}) we used the sum rule and electron hole
symmetry, Eqs.~(C.1) and (C.6) of Ref.~\cite{rDatta1}. We distinguish
between $\tilde{N}^S_{Sh,out}(E)$ and $\tilde{N}^S_{Se,in}(E)$ in
Eq.~(\ref{I1s}), since the incoming and outgoing electron-like
quasi-particles have different densities of states. The second term in
Eq.~(\ref{defJQ/JP}) results in a small correction current $I_2$,
proportional to $(q/k_F)$. The rather cumbersome expression for $I_2$
is
\begin{eqnarray}
I_2 & = & \frac{2e}{h} \int_{-\infty}^{\infty}  
[(q/k)(|u_{kq}|^2 - |v_{kq}|^2)]_{Se,out} 
\tilde{N}^S_{Se,out}(E) T_{Se,Ne}(E)
[f(E-eV) - f(E)] dE
\nonumber \\
& - & \frac{2e}{h} \int_{-\infty}^{\infty}  
[(q/k)(|u_{kq}|^2 - |v_{kq}|^2)]_{Sh,out}
\tilde{N}^S_{Sh,out}(E) T_{Sh,Ne}(E)] 
[f(E-eV) - f(E)] dE
\nonumber \\
& + & \frac{2e}{h} \int_{-\infty}^{\infty} 
[(q/k)(|u_{kq}|^2 - |v_{kq}|^2)]_{Se,out}
\tilde{N}^S_{Se,out}(E) [2f(E)-1] dE
\nonumber \\
& - & \frac{2e}{h} \int_{-\infty}^{\infty} 
[(q/k)(|u_{kq}|^2 - |v_{kq}|^2)]_{Se,in}
\tilde{N}^S_{Se,in}(E) [2f(E)-1] dE .
\label{I2s}
\end{eqnarray}

We again translate between $k$ and $E$ inside the superconductor using
Eq.~(\ref{dispersion}), taking care to assign the appropriate branch
of the dispersion curve for incoming or outgoing electron- or hole-
like particles. The first two terms in $I_2$ are a small correction to the
superfluid flow due to additional quasi-particle injection, while the
last two terms are a small correction to the equilibrium superfluid flow.
A rigorous treatment guarenteeing global current conservation at
any temperature would equate the current $I$ from Eq.~(\ref{I_N})
with the current $I_S$ from Eq.~(\ref{IS}), adjusting the superfluid
velocity $q$ until $I = I_S$. 
} 

\twocolumn

We now analyze the validity of our `two-fluid'' procedure from Section
\ref{secwhymore}. We henceforth neglect the current $I_2$ as
insignificant compared with $I_1$.  At zero temperature, the current
operator evaluated inside the superconductor we find from from
Eq.~(\ref{I1s}) as
\begin{eqnarray}
I_1(T=0) & = & \frac{2e}{h} \int_{0}^{eV} 
[\tilde{N}^S_{Se,out}(E) T_{Se,Ne}(E) \nonumber \\
& - & \tilde{N}^S_{Sh,out}(E) T_{Sh,Ne}(E)] dE \nonumber \\
& + & \frac{4e\Delta}{h} (v_s/v_d).
\label{I1sT0}
\end{eqnarray}
The zero temperature limit of Eq.~(\ref{I_N}) is given in Section 
\ref{secwhymore} as Eqs.~(\ref{I_QP})-(\ref{2And}). The second term
in Eqs.(\ref{I1sT0}) is the superfluid flow term $I_C$. So we can
certainly identify $I_A$ from Eq.~(\ref{2And}) with $I_C$.
However, Eq.~(\ref{I_QP}) for $I_{QP}$ is not exactly equal to
the first term in Eq.~(\ref{I1sT0}), the difference being the
additional factor of the superconducting density of states
$\tilde{N}^S(E)$ in Eq.~(\ref{I1sT0}). 

For the Cooper pair flow away from the NS interface (which we are
considering in this paper), and for the energy range between $0$ and
$eV$, the outgoing hole-like quasi-particle conduction channel opens
at a lower energy than the electron-like quasi-particle channel (see
Fig.~\ref{figBand}). This
means the first term in Eq.~(\ref{I1sT0}) will be larger and more
negative than $I_{QP}$ from Eq.~(\ref{2And}), requiring a larger value
of the superfluid velocity $v_s$ at each value of the applied voltage
$V$ than in the two-fluid model of Section \ref{secwhymore}. We
conclude that the treatment in Section \ref{secwhymore} therefore
underestimates the effect of superfluid flow on the excess current.

\end{document}